\documentclass[ prl, onecolumn, superscriptaddress]{revtex4-1}
\usepackage{amssymb, graphics, epsfig, color, ulem, float,graphicx}
\usepackage[english]{babel}
\usepackage{amsmath,amstext,bm,verbatim,bbold}
\usepackage{graphicx,subfigure}
\usepackage{epstopdf}
\usepackage[amssymb]{SIunits}
\usepackage[english,pdfauthor={Marcel Mokbel},  pdfauthor={Marcel Mokbel}, pdftitle={Paper}, breaklinks=false, colorlinks=false, linkbordercolor={0 0 1}, linktocpage=true, hyperfootnotes=true]{hyperref}
\usepackage{todonotes}

\newcommand{\tss}[1]{$\scriptstyle^{#1}$}

\usepackage[all]{hypcap} 

\addto\extrasenglish{%
}

\begin{document}

\title{The Poisson ratio of the cellular actin cortex is frequency-dependent}
\author{M. Mokbel}
\affiliation{Faculty of Informatics/Mathematics, Hochschule f\"ur Technik und Wirtschaft, Dresden, Germany}
\author{K. Hosseini}
\affiliation{Biotechnology Center, Technische Universit\"at Dresden, Dresden, Germany}
\affiliation{Cluster of Excellence Physics of Life, Technische Universit\"at Dresden, Dresden, Germany}
\author{S. Aland}
\email{Corresponding author: sebastian.aland@htw-dresden.de}
\affiliation{Faculty of Informatics/Mathematics, Hochschule f\"ur Technik und Wirtschaft, Dresden, Germany}
\author{E. Fischer-Friedrich}
\email{Corresponding author: elisabeth.fischer-friedrich@tu-dresden.de}
\affiliation{Biotechnology Center, Technische Universit\"at Dresden, Dresden, Germany}
\affiliation{Cluster of Excellence Physics of Life, Technische Universit\"at Dresden, Dresden, Germany}

\begin{abstract}{
Cell shape changes are vital for many physiological processes such as cell proliferation, cell migration and morphogenesis. They emerge from an orchestrated interplay of active cellular force generation and passive cellular force response - both crucially influenced by the actin cytoskeleton. 
To model cellular force response and deformation, cell mechanical models commonly describe the actin cytoskeleton as a contractile isotropic incompressible material. However, in particular at slow frequencies, there is no compelling reason to assume incompressibility as the water content of the cytoskeleton may change.
Here we challenge the assumption of incompressibility by comparing computer simulations of an isotropic actin cortex with tunable Poisson ratio  to measured cellular force response. 
Comparing simulation results and experimental data, we determine the Poisson ratio of the cortex in a frequency-dependent manner.
We find that the Poisson ratio of the cortex decreases with frequency likely due to actin cortex turnover leading to an over-proportional decrease of shear stiffness at larger time scales.
We thus report a trend of the Poisson ratio similar to that of glassy materials, where the frequency-dependence of jamming leads to an analogous effect.
}
\end{abstract}
\maketitle


\section{ Introduction}
The actin cytoskeleton, a cross-linked meshwork of actin polymers, is a key structural element that crucially influences mechanical properties of cells \cite{salb12}. In fact, for rounded mitotic cells, the mitotic actin cortex, a thin actin cytoskeleton layer attached to the plasma membrane,  could be shown to be the dominant mechanical structure in whole-cell deformations \cite{fisc16}.\\
In the past, cell mechanical models have been developed to rationalize cell deformation in different biological systems \cite{pull07, koll11}.
Commonly, these models describe the actin cytoskeleton as a contractile isotropic incompressible material  \cite{juli07}. 
The assumption of incompressibility implies a Poisson ratio of $0.5$. Incompressibility of the actin cytoskeleton is motivated by incompressibility of water and high water content in the actin cytoskeleton \cite{dimi02}. 
This assumption is justified for high-frequency deformations as in this case substantial water movement past the elastic scaffold of the polymerized actin meshwork would give rise to strong friction and is thus energetically suppressed (see Supplementary Section~1). The anticipated high-frequency incompressibility was confirmed experimentally in {\it in vitro} reconstituted actin meshworks in a frequency range of $~500-10,000\,$Hz  \cite{koen06}. 
However, in particular at slow frequencies, there is no compelling reason to assume incompressibility as the water content of the cytoskeleton may change via water fluxes past the cytoskeletal scaffold leading to a bulk compression or dilation. Furthermore, the actin cytoskeleton is subject to dynamic turnover \cite{salb12} and exhibits viscoelastic material properties \cite{fabr01, fisc16, koll11, pull07}. Therefore, it is expected that the cortical Poisson ratio is frequency-dependent as has been reported for other viscoelastic materials such as acrylic glass. There, the Poisson ratio was shown to increase from ~0.32 to 0.5 for increasing time scales \cite{grea11,lu97}. 
\\
Here we critically examine the assumption of actin cortex incompressibility by measuring the Poisson ratio of the actin cortex in dependence of the frequency of time-periodic deformations.
To this end, we compare the  measured force response of the actin cortex in HeLa cells in mitotic arrest to the simulated force response of elastic model cortices with known Poisson ratio. 
\\
\begin{figure}
\centering
\includegraphics[width=12cm]{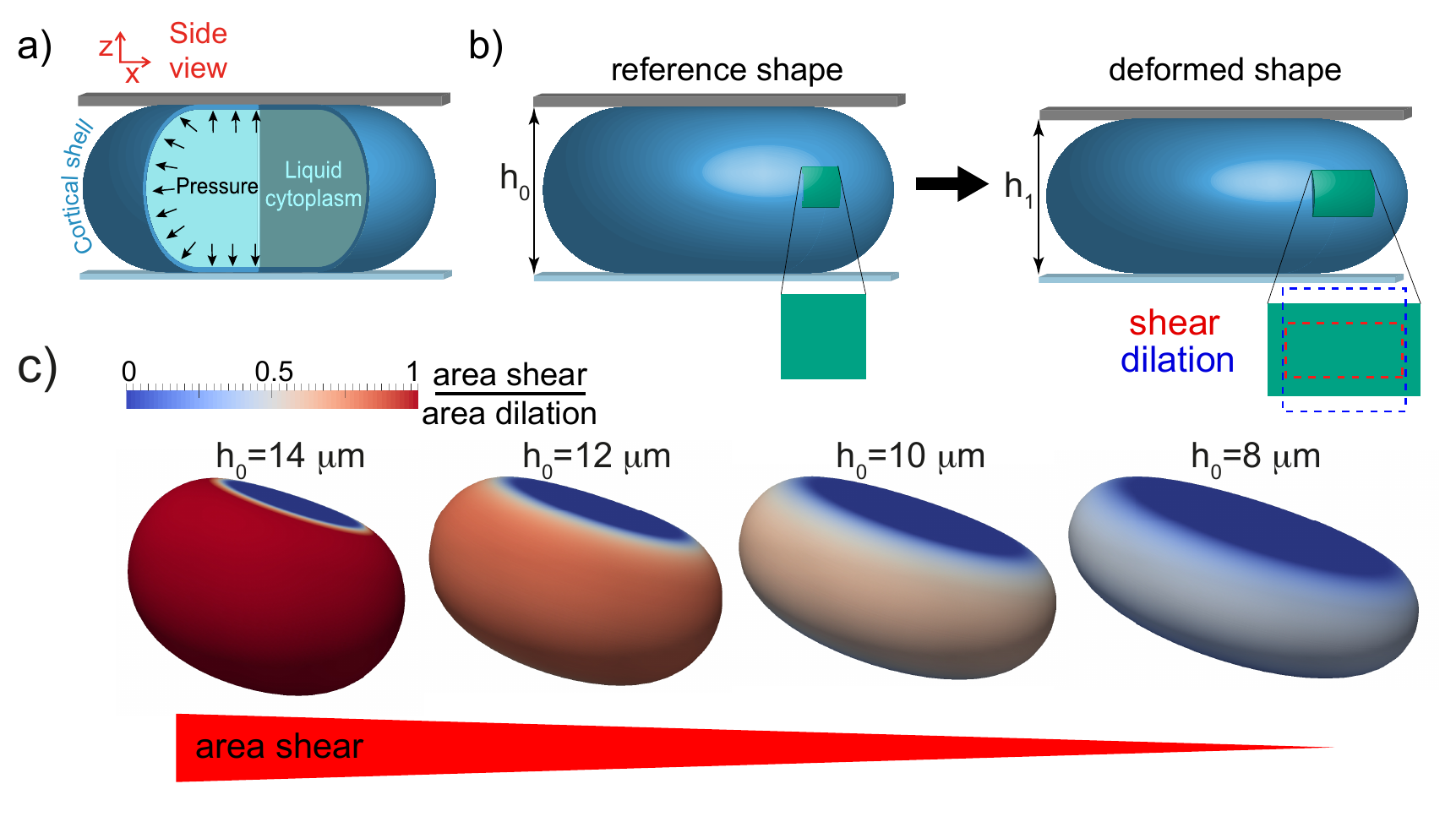}
\caption{ \label{fig:Fig1}
Elastic uniaxial compression of a cortical shell.
a) Cell-mechanical model.
b) Left panel: a square-shaped surface element (green) in the elastic reference shape of the shell.  
Right panel: after a small amount of uniaxial compression through reduction of shell height, the surface element is deformed (deformation is exaggerrated here for illustration purposes). 
c) Elastic deformation of model cells  exhibit a decreasing ratio of area shear to area dilation at decreasing reference cell heights (simulation parameters as in \autoref{fig:Fig2}).
}
\end{figure}
\section{Theory of cortical shell deformation}
We numerically determine the mechanical response to uniaxial compression steps of idealized model cells. Our model cells are constituted by an isotropic contractile  elastic thin shell mimicking the actin cortex  \cite{clar13}. This shell encloses an incompressible liquid interior representing the cytoplasm. 
Cortical shells are thus assumed to enclose a constant volume $V_{cell}$ independent of elastic stresses as the associated hydrostatic pressures in the cell are negligible as compared to the osmotic pressure of the medium \cite{clar11}. 
We assume a model shell thickness $t_c$ of $200\,$nm as measured before for the actin cortex of mitotic HeLa cells \cite{clar13}, and a model cell volume of $V_{cell}=4300\,\mu{\rm m^3}$ which was approximately the average volume of mitotic HeLa cells in our experiments. 
According to elasticity theory, the shell's elastic behavior is characterized by three elastic moduli - i) the area bulk modulus  $K_B$ characterising the  resistance to area dilation or compression, ii) the area shear modulus $K_S$ characterizing the  resistance to shear deformation of a surface patch of the shell, and iii) the bending modulus $B$ characterizing the resistance to shell bending. In the case of an isotropic material, only two of the three moduli are independent and we have $K_B=t_c E/(2(1-\nu)), K_S=t_c E/(2(1+\nu))$ and $B=t_c^3 E/(24(1-\nu^2))$, where $E$ is the Young's modulus of the shell.

We consider model cells that are confined between two parallel plates in an elastic reference configuration of height $h_0$, see \autoref{fig:Fig1}(a,b).
There, we anticipate a constant isotropic contractile in-plane stress $\sigma_a$ in the cortical shell that captures active actomyosin contractility of the actin cortex which gives rise to a constant active cortical tension $\gamma_a= t_c \sigma_a$. This active tension is balanced by the internal hydrostatic pressure of the liquid interior. In the absence of elastic stresses, the contractile tension $\gamma_a$ drives the model cell into the shape of a non-adherent droplet with constant mean curvature $H$ in the regions of unsupported shell surface \cite{fisc14}.
We use these confined droplet shapes as elastic reference configuration since the actin cortex has been previously characterized to be viscoelastic with complete stress relaxations after $\le 1$ minute \cite {fisc16}. Therefore, mechanically confining cells to a  height $h$ leads to a new droplet-shaped reference shape of height $h$ after a short waiting time.
In this elastic reference state, a model cell exerts a constant force due to active tension 
\begin{equation}
F_a(h)= 2\gamma_a(h) H(h)\, A_{c}(h) \label{eq:Eq1}
\end{equation}
on the confining plates, where $A_{c}(h)$ is the circular contact area between the cell and the plate and $H$ is the mean curvature, both at height $h$ of the cell \cite{fisc14, fisc16}. \\
This force exerted on the confining plates is the central quantity of our investigations as we can measure it in our experiments and compute it in our Finite-Element simulations \cite{Mokbel}.
To probe the force response of a model cell, steps of uniaxial compression  are imposed that lower the cell height from a starting height $h_0$ to $h_1=h_0-\Delta h$. In turn, the shell material is deformed and elastic stresses are induced (\autoref{fig:Fig1}(a,b)). Together with an increase of the shell's plate contact, this contributes to an increase of the force exerted on the confining plates. 
The new force for the decreased plate distance $h_1$ is denoted as $F_{tot}(h_0, \Delta h)=F_a(h_1)+\Delta F(h_0, \Delta h)$,
where $\Delta F(h_0, \Delta h)$ captures the elastic contribution of the force increase. For our study, we consider small compression steps where $\Delta F(h_0, \Delta h)$ is well approximated as a linear function of $\Delta h$.
Further, we verified that the force response of the liquid interior adds $\le 1\%$ to the effective modulus for cytoplasmic viscosities of up to $1\,$Pa$\cdot$s, oscillation frequencies $\le \,10\,$Hz (see Supplementary Section~2) and relative cell confinement lower than $80\%$.
Therefore, we henceforth neglect viscous flows in the cytoplasm simulating only the elastic deformation of a shell and an internal pressure.\\
In analogy to Eq.~\eqref{eq:Eq1}, we can relate the overall force of the cortex after elastic deformation to an effective cortical tension \cite{fisc16}
\begin{equation}
\gamma_{eff}(h_0, \Delta h)= \frac{F_{tot}(h_0, \Delta h)}{2H(h_1) A_{c}(h_1)} \label{eq:EffTens},
\end{equation}
where $\gamma_{eff}=\gamma_a+ \Delta \gamma_{eff}$ with $\Delta \gamma_{eff}=\Delta F(h_0,\Delta h)/(2H(h_1) A_{c}(h_1))$.
From simulation results, we determine an effective elastic modulus as 
\begin{equation}
K(h)= \frac{\Delta \gamma_{eff}}{\epsilon_A}  \label{eq:EffElasMod},
\end{equation} 
where $h=(h_0+h_1)/2$ and $\epsilon_A$ is the surface area strain 
\begin{equation}
\epsilon_A=\Delta A(h_0, \Delta h)/A(h_0)  \label{eq:AreaStrain}
\end{equation}
 with $\Delta A$  the increase in overall surface area of the model cell through deformation and $A(h_0)$ the original surface area at height $h_0$ in the absence of elastic stresses  \cite{fisc16}. We estimate the geometrical parameters  $\Delta A$,  $H$ and $A_{c}$, assuming a droplet shape of the cell before and after deformation \cite{fisc14}. We verified that the estimated surface increase $\Delta A$ deviates less than 6 \% from the area increase calculated in simulations providing thus a good approximation. 

\begin{figure}
\centering
\includegraphics[width=12cm]{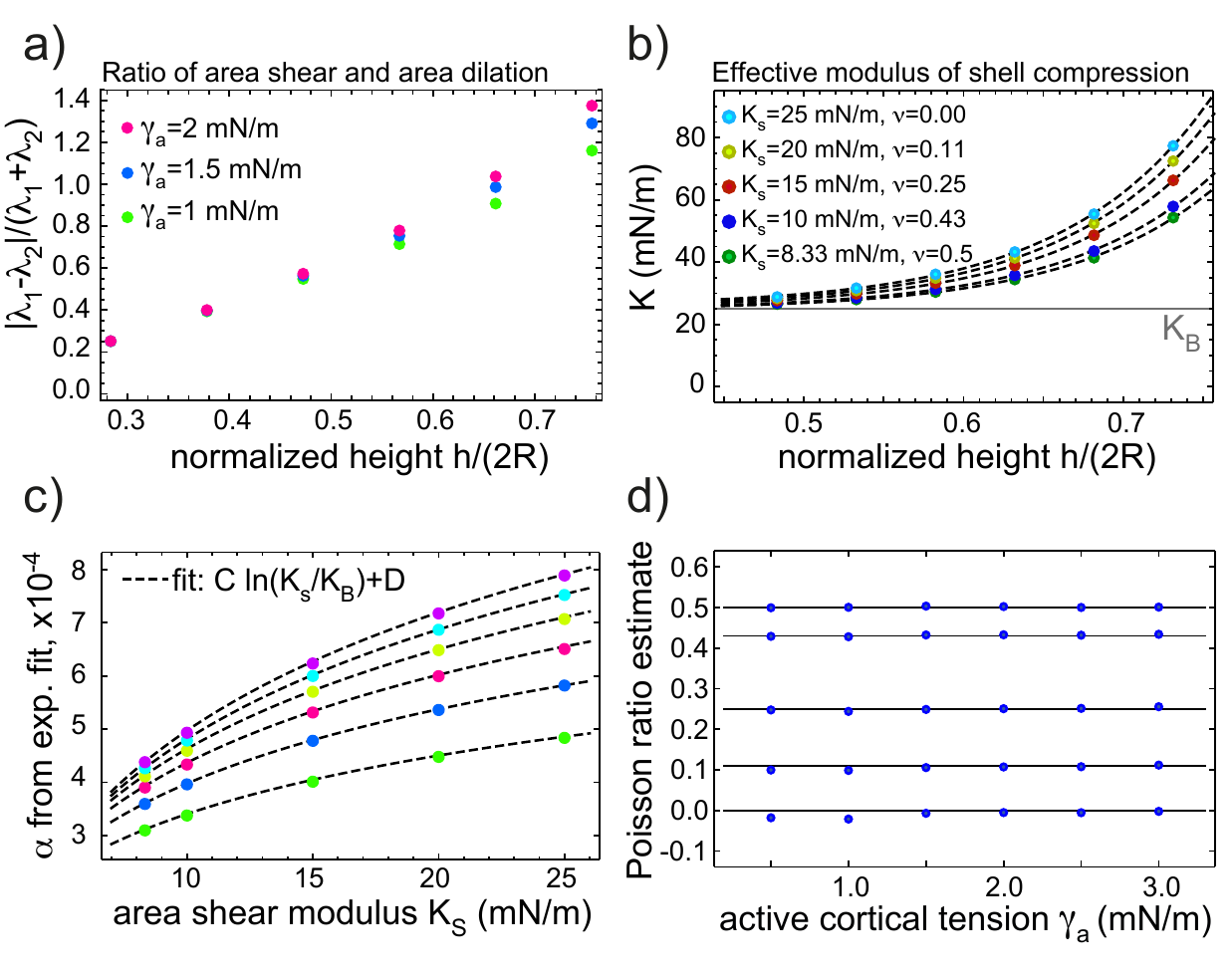}
\caption{ \label{fig:Fig2}
Uniaxial compression of elastic model cells with varying reference height. 
a) Ratio of  area shear  to area dilation at the shell equator quantified as $|\lambda_1-\lambda_2|/(\lambda_1+\lambda_2)$, where $\lambda_1$ and $\lambda_2$ are the equatorial eigenvalues of the in-plane shear tensor.
b) Effective elastic modulus $K$  in dependence of mean shell height $h=h_0-\Delta h/2$  (dashed lines: fit by Eq.~\eqref{eq:ExpLaw}). 
c) Fit coefficient $\alpha$ in dependence of  $K_S$ for varying cortical tensions $\gamma_a$ (bottom to top: $0.5-3\,{\rm mN/m}$ in increments of $0.5{\rm mN/m}$, dashed lines: fit by Eq.~\eqref{eq:LogLaw}). 
The choice of cortical tension reflects the range of experimental values.
d) Reconstructed Poisson ratios (blue dots) from effective elastic moduli as shown in panel b. Black lines indicate actual Poisson ratio values of underlying simulations.
Elastic parameters were chosen to be $K_B=25\,$mN/m and $K_S=8.3, 10, 15, 20$ or $25\,$mN/m corresponding to Poisson ratios of $\nu=0.5, 0.43, 0.25, 0.11$ and $0$. Values of $K_B$ were motivated by measurement results reported by \cite{fisc16}. The cell volume was $4300\,\mu{\rm m}^3$ and $\Delta h=0.5\,\mu$m.
}
\end{figure}

Finite-Element simulations were carried out to extract the effective elastic modulus $K$ for 540 combinations of cell heights, area shear moduli, bending stiffnesses and surface tensions (see Supplementary Section~3 and 4). For convenience, we introduce now the normalised cell height $\tilde h= h/(2 R_{cell})$ with $R_{cell}=(\frac{3}{4\pi}  V_{cell})^{1/3}$. 
We find that at low values of normalized reference cell height $\tilde h$,  the effective modulus $K$ approaches the area bulk modulus $K_B$ due to dominance of area dilation over area shear during shell deformation (\autoref{fig:Fig2}(a,b)). For larger normalized heights $\tilde h$, the effective modulus $K$ increases  due to an increasing contribution of area shear during model cell deformation (\autoref{fig:Fig2}(b)). We can capture this increase phenomenologically by an exponential rise
\begin{equation}
K(\tilde h)\approx K_B(1+\alpha \exp(\tilde h/\lambda)), \label{eq:ExpLaw}
\end{equation}
where $\lambda\approx 0.09$ (dashed lines in \autoref{fig:Fig2}(b), see Supplementary Section~3).
The amplitude of the exponential increase $\alpha$ depends on the normalized shear modulus $\tilde K_S=K_S/K_B$ as well as the normalized surface tension $\tilde \gamma_a=\gamma_a/K_B$. In the experimentally relevant range $0.45<\tilde h<0.75$, we capture this dependence again by a phenomenological law 
\begin{align}
\alpha(\tilde K_S, \tilde \gamma_a)&\approx C(\tilde\gamma_a) \log[\tilde K_S] + D(\tilde\gamma_a) \label{eq:LogLaw}
\end{align}
where $C(\tilde\gamma_a)$ and $D(\tilde\gamma_a)$ are polynomials of third degree in $\tilde \gamma_a$ (dashed lines in \autoref{fig:Fig2}(c), Supplementary Section~3).\\
Eqn.~\eqref{eq:ExpLaw} and \eqref{eq:LogLaw} provide  now an analysis scheme to  reconstruct the Poisson ratio from measured effective moduli $K(\tilde h)$ for known $\gamma_a$ (\autoref{fig:Fig2}(d)).\\
\section{ Experimental results}
\begin{figure}[h]
\centering
\includegraphics[width=12cm]{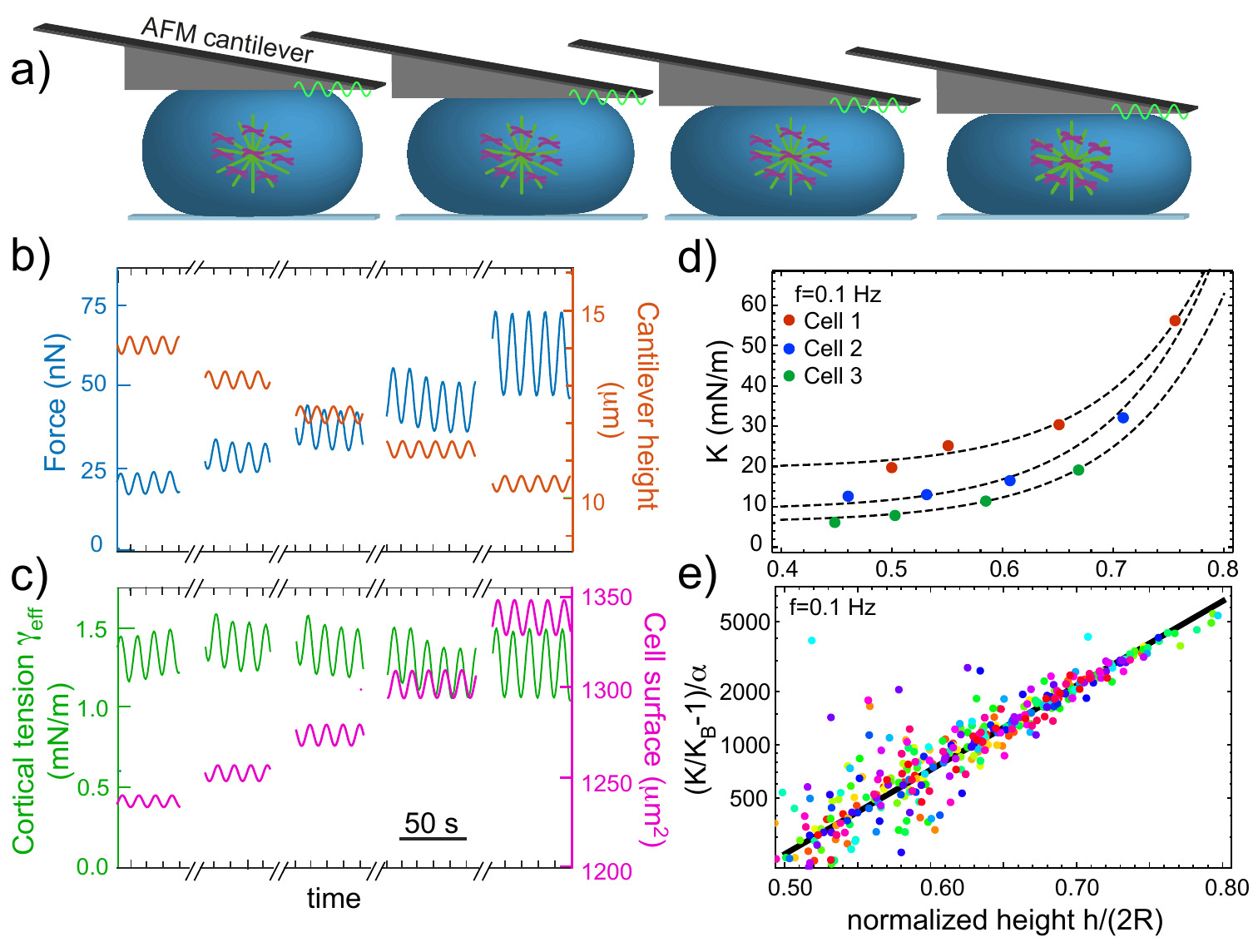}
\label{fig:Fig3}
\caption{ 
AFM-based deformation of HeLa cells. 
a) Cells in mitotic arrest were confined through  a wedged cantilever (schematic, green: microtubules, violet: chromosomes). Oscillatory cell height modulations are applied at decreasing mean cell heights.
b) Exemplary force and cantilever height output  at $f=0.1\,$Hz.
c) Values of cortical tension and cell surface area associated to panel a calculated from force and cantilever height as described before in  \cite{fisc16}.
d) Exemplary effective elastic moduli of cell cortices versus normalized cell heights. Dashed lines show a fit according to Eq.~\eqref{eq:ExpLaw} with fit parameters $K_B$ and $\alpha$.
e) Normalized effective elastic moduli $(K/K_B -1)/\alpha$  of all cells measured at $f=0.1\,$Hz.
The phenomenological dependence predicted by Eq.~\eqref{eq:ExpLaw} is captured by the solid black line. Different colors represent different cells. 
}
\end{figure}

\begin{figure}[h]
\centering
\includegraphics[width=12cm]{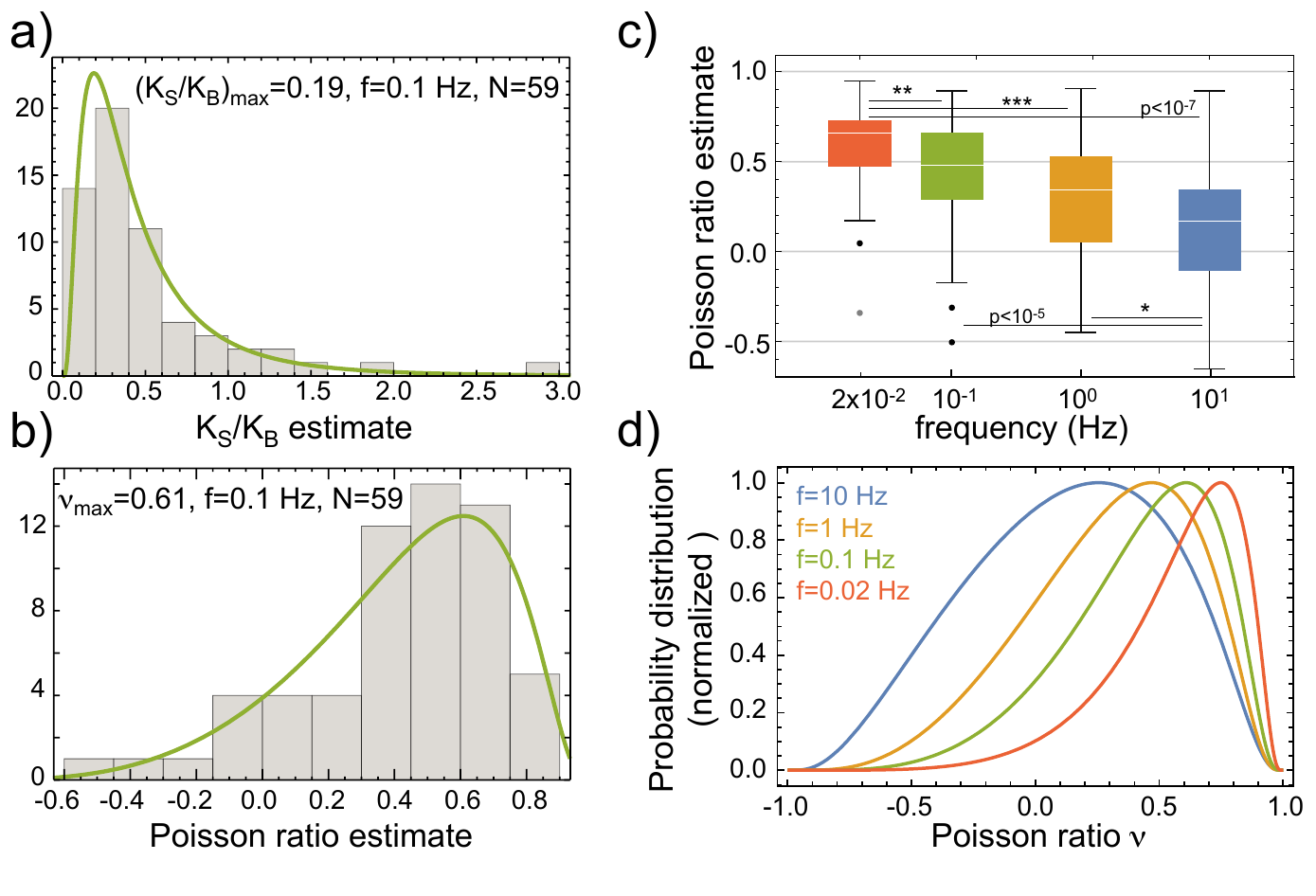}\label{fig:Fig4}
\caption{
Poisson ratio estimates of the actin cortex in mitotic HeLa cells.
a) Histogram of estimated  $K_S/K_B$  at $f=0.1\,$Hz (green line represents  lognormal distribution of maximum likelihood).
b) Histogram of corresponding Poisson ratios (green line: distribution induced by lognormal distribution in a).
c) Box-plots of estimated Poisson ratios at different frequencies. From left to right, median values: 0.66, 0.48, 0.34, 0.17, IQR: 0.26, 0.38, 0.48, 0.45.
d) Fitted distributions of estimated Poisson ratios for different measured frequencies (analogous to solid line in e, bottom). 
}
\end{figure}

We now want to use our theoretical insight to determine the Poisson ratio of the actin cortex in live cells. As a cellular model system, we use HeLa cells in mitotic arrest since they are void of a nucleus and exhibit a large cell surface tension that ensures droplet-shaped cells in confinement \cite{fisc14}. We mechanically deform these cells in an oscillatory manner around different heights of confinement via the wedged cantilever of an atomic force microscope  (\autoref{fig:Fig3}(a)) \cite{fisc14, fisc16}. During these measurements, we record the force exerted by the AFM cantilever and the respective cantilever height $h_{cant}$ (\autoref{fig:Fig3}(b)). We then calculate the associated time-periodic  effective cortical tension $\gamma_{eff}(t)$ and area strain $\epsilon_A(t)$ according to Eq.~\eqref{eq:EffTens} and Eq.~\eqref{eq:AreaStrain} with $h_1(t)=h_{cant}(t)$, $ h= <h_{cant}(t)>$ and $\Delta h(t)=h_1(t)-h$ (\autoref{fig:Fig3}(c)). We determine the volume of the measured cell $V_{cell}$ from imaging (see Materials and Methods in Supplementary Section~5) and calculate an associated cell radius $R_{cell}$.  In analogy to Eq.~\eqref{eq:EffElasMod}, we infer an effective modulus of the actin cortex of measured cells 
$K= \frac{{\hat\gamma_{eff}}}{{\hat \epsilon_A}}$,
where ${\hat\gamma_{eff}}$ and ${\hat\epsilon_A}$ are the amplitudes of the time-periodic signal of $\gamma_{eff}$ and $\epsilon_A=\Delta A/<A>$, respectively (\autoref{fig:Fig3}(d))\cite{fisc16}. 
Our measurement and analysis procedure is repeated at different cell heights to obtain $K$ as a function of normalised cell confinement height $\tilde h$ (\autoref{fig:Fig3}(d)).
\\
%
Cell-mechanical measurements are performed at frequencies $0.02\,$, $0.1\,$, $1\,$ and $10\,$Hz. 
Using the correspondence principle, we  apply our insight on the mechanical response of elastic model cells to our measurements of viscoelastic live cells \cite{lake17}:
we fit the measured cortical modulus $K$ in dependence of cell height by  Eq.~\eqref{eq:ExpLaw} and obtain the fit parameter $\alpha$ and $K_B$ (\autoref{fig:Fig3}(e)). In general, we find a good agreement between measured values and the exponential increase predicted by our elastic shell calculations with a median r-squared value of $~0.94$  for $f=0.1-10\,$Hz and 0.84 for $f=0.02\,$Hz. 
The good agreement between data points and the fitting function provided by numerical simulation illustrates the suitability of our cell-mechanical description.\\
Furthermore, we estimate the cortical tension as the time-average $\gamma_a\approx <\gamma_{eff}>$. Inverting Eq.~\eqref{eq:LogLaw}, we obtain an estimate for $\tilde K_S=K_S/K_B$ and thus the Poisson ratio $\nu$ (\autoref{fig:Fig4}(a,b,c), Supplementary Fig.~3(a,b)). We find, that the obtained Poisson ratio estimate depends on the frequency of time-periodic cell deformations with lower Poisson ratios for fast cell deformations. Median values of the Poisson ratio vary between values of  $0.17$ and $ 0.48$ for decreasing frequencies between $10-0.1\,$Hz (\autoref{fig:Fig4}(c,d)). For the slowest frequency $0.02\,$Hz, where cortex turnover is expected to influence cell mechanics, we estimate a median Poisson ratio of $0.66$ (\autoref{fig:Fig4}(c,d)). 
%
Our results show a substantial scatter of Poisson ratio estimates at a given frequency (\autoref{fig:Fig4}(c)). In order to examine the origin of this statistical spread, we quantify the influence of experimental uncertainties. To this end, we access the error of our cell volume estimate to be $~7.5\%$ and of cell height to be $~0.5\,\mu$m.
In turn, we calculate the resulting variation of Poisson ratios for elastic model cells with a known Poisson ratio by introducing corresponding artificial errors in cell volume and cell height (see Supplementary Fig.~3(c)). In this way, we find resulting interquartile ranges (IQR) between $0.24$ and $0.39$, which are close to IQR values found for experimental spreads. Therefore, we conclude that statistical scatter in our experimental data stems to a substantial amount from measurement errors and not exclusively from cell-cell variations.
Amongst cell-cell variations, we expect variations in cortical thickness, and thus in the contribution of bending stiffness to cell deformations, as a major source for variations in Poisson ratio estimates  (see Supplementary Section~7). 
%
In summary, in spite of large statistical scatter, we observe a robust, significant trend of increasing Poisson ratio values of the mitotic actin cytoskeleton with decreasing frequency (\autoref{fig:Fig4}(c,d)) where Poisson ratio distributions are significantly different from each other at different frequencies (two-sided Mann-Whitney test: p-values $\leq 0.02$ for neighbouring frequencies, $\leq 10^{-4}$ for all other frequency pairs).\\
\section{ Discussion}
Here, we report  a new measurement method to determine the Poisson ratio of the actin cortex in biological cells that is based on the time-periodic  deformation of initially round mitotic cells through the wedged cantilever of an atomic force microscope. The key idea behind this technique is that mechanical deformation at different reference shapes probes the cortical shell at varying contributions of  area dilation and area shear (\autoref{fig:Fig1}(b,c) and \autoref{fig:Fig2}(a)).  \\
For our measurements results at the largest frequency ($f=10\,$Hz), we expect that cortex turnover plays a negligible role for the mechanical properties of the cortex \cite{salb12}. There, we find a median Poisson ratio of $0.17$.
This value is considerably lower than the incompressible case of $\nu=0.5$ and reasonably close to theoretical predictions of $0.25$ for foamed elastic materials or polymer gels \cite{gent59, geis80}. \\
Furthermore, we find a clear trend for the Poisson ratio to increase with time scale; median values of the Poisson ratio increase from  $0.17$  to $0.66$ in a time scale range of $\tau\approx 0.016-8\,$s associated to a frequency range of $f=0.02-10\,$Hz by $\tau=1/(2\pi f)$ (\autoref{fig:Fig4}(c,d)).  A plausible explanation for this trend is that turnover of actin cross-linkers (taking place on time scales of $\sim 0.2-20\,$s \cite{salb12}) leads to a significant decrease of the shear modulus at growing time scales but to a minor change of the bulk modulus of the actin cortex. Correspondingly, the Poisson ratio would decrease with time scale and increase with frequency.
In fact, a similar effect was reported as a hallmark for the glass transition of synthetic polymer materials  \cite{grea11}. There, an increase of Poisson ratio as a function of time scale was reported when moving from glassy to rubbery rheological behaviour. Correspondingly, this transition is accompanied by a strong decrease of the shear modulus due to jamming release but a minor decrease of the bulk modulus with time scale \cite{grea11, tsch02, lu97}. \\
It is noteworthy that for a thin shell of an isotropic material the associated two-dimensional Poisson ratio $\nu_{2d}$ coincides with the three-dimensional Poisson ratio $\nu$. In this case,  $\nu_{2d}$ may adopt values in the range $[-1, 0.5]$ (see Supplementary Section 8).  However, if the assumption of material isotropy is relaxed, $\nu_{2d}$ may adopt values that may reach up to $1$. For the slowest frequency probed in our measurements, the Poisson ratio estimate exceeds $0.5$. This might hint at a violation of cortical isotropy  at slower frequencies. 
Cortical turnover is critically influenced by the cortex interface with the plasma membrane \cite{char06, salb12} which might account for the emergence of anisotropy at large time scales. \\
Poisson ratios of cellular material have been previously estimated:
Mahaffy \textit{et al.} developed a method to estimate the Poisson ratio of adherent cells through slow AFM indentation at a gradually increasing indentation depth into a thin cytoskeletal layer above a substrate \cite{maha04}. Poisson ratio estimates from this method are between $0.4-0.5$  \cite{maha04,betz11, lu13}. 
Trickey \textit{et al.} \cite{tric06} measured the Poisson ratio of chondrocytes through a whole-cell perturbation via micropipette aspiration and subsequent shape relaxation thereby estimating values of $0.38$. 
However, both methods \cite{maha04, tric06} ignored the possible time-scale dependence of the Poisson ratio.
This fact makes it hard to compare these earlier findings to our data. We do, however, anticipate that our measurement results do not contradict with those previous measurements due to our comparable results in the frequency range $0.1-1\,$Hz.\\
For {\it in vitro} reconstituted branched actin meshworks, Bussonnier {\it et al.} clearly showed compressibility of branched actin meshworks on a time scale of few seconds (Poisson ratio between 0.1-0.2) \cite{buss14}.  
By contrast, entangled actin meshworks without cross-linking were shown to be close to imcompressible \cite{gard03}. 
This discrepancy indicates that not only the time-scale but also the presence of actin cross-linkers plays a crucial role for the Poisson ratio of actin meshworks.
 \\
To our best knowledge, we present here for the first time measurements of the Poisson ratio of the actin cortex in live cells in dependence of frequency showing a clear frequency-dependent trend. Therefore, we give evidence that the actin cortex may not in general be treated as an incompressible material. 

\section*{Acknowledgments}
We thank Jochen Guck, Isabel Richter and Anna Taubenberger for access and introduction to infrastructure in the lab. In addition, we thank the CMCB light microscopy facility for excellent support.
SA acknowledges support from the German Science Foundation (grant AL 1705/3) and tax money based on the
budget passed by the delegates of the Saxonian state parliament.
EFF thanks for financial support from the DFG, project FI 2260/4-1. 
SA and EFF acknowledge financial support from the DFG in the context of the Forschergruppe FOR3013, projects AL 1705/6-1 (SA) and FI 2260/5-1 (EFF).
Simulations were performed at the Center for Information Services and High Performance Computing (ZIH) at TU Dresden.

\section*{Author Contributions}
S.A. and E.F.F. designed the research. M.M. and S.A. developed the numerical method. M.M. performed simulations. K.H. and E.F.F. performed the experiments. M.M., K.H. and E.F.F. performed data analysis. M.M., S.A. and E.F.F. wrote the manuscript.

\section*{Competing Interests}
The authors declare no competing interests.


\newpage
\part{Supplementary Material}

\section{Effective compression modulus of a thin poroelastic layer}
\label{sec:AppSec1}
For a poroelastic material consisting of a viscoelastic porous scaffold and an immersing fluid, we have in Cartesian coordinates\cite{biot57}
\begin{equation}
\epsilon_{ii}=\frac{1}{K}\left(\frac{\sigma_{ii}}{3}+ p_H\right), \label{eq:Poro1}
\end{equation}
where $ p_H$ is the hydrostatic pressure increment in the fluid, $K$ is the bulk modulus of the scaffold material, and $\sigma_{ij}$  and $\epsilon_{ij}$ are the components of the stress and strain tensor of the elastic scaffold, respectively. Using Darcy's law, one obtains \cite{biot57}
\begin{equation}
\frac{k_{perm}}{\eta} \Delta p_H =\frac{\partial\epsilon_{ii}}{\partial t}, \label{eq:Poro2}
\end{equation}
where $k_{perm}$ characterizes the permeability of the scaffold material, $\eta$ is the viscosity of the immersing fluid  and $\Delta$ is the Laplace operator. 
Consider a flat horizontal layer of porous material with thickness  $t_c$. We choose the middle layer of the layer to be at  coordinate $z=0$. 
Consider that oscillating opposing uniform forces are applied at the top and the bottom of the layer by a porous slab such that a small time-periodic (sinusoidal) compression  is achieved. The edges of the layer are clamped such that displacement in x- and y-direction are prohibited. In this case, the trace of the strain tensor is $\epsilon_{ii}=\epsilon_{zz}$. Equivalently, $\sigma_{ii}=\sigma_{zz}$, where $\sigma_{zz}$ varies time-periodically but is spatially uniform due to the force balance requirement $\partial_z \sigma_{zz}=0$. 
According to Eqn.~\eqref{eq:Poro1} and \eqref{eq:Poro2}, we have $\epsilon_{zz}=\frac{1}{K}(\frac{\sigma_{zz}}{3}+p_H)$ and $i\omega \epsilon_{zz}=\frac{k_{perm}}{\eta}\partial_z^2 p_H$, where we identified the time-derivative with a multiplication by $i\omega$.
We therefore obtain the following partial differential equation in $p_H$
\begin{equation}
\frac{i\omega }{K}\left(\frac{\sigma_{zz}}{3}+p_H\right)=\frac{k_{perm}}{\eta}\partial^2_z p_H \label{eq:Poro3}.
\end{equation}
A special solution of Eq.~\eqref{eq:Poro3} is $p_H=-\sigma_{zz}/3$. The general solution of the corrsponding homogeneous equation reads $p_H^h(z)=A e^{\frac{z}{\lambda}}+Be^{-\frac{z}{\lambda}}$, where $\lambda=\sqrt{\frac{Kk_{perm}}{i\omega\eta}}$. Assuming that the porosity of the confining slabs is significantly larger than the porosity of the poroelastic layer, we impose the boundary conditions\cite{biot57} $p_H(z=\pm t_c/2)=0$ and obtain the full solution
\begin{equation}
p_H(t,z)=\frac{\sigma_{zz}(t)}{3} \left(\frac{(e^{\frac{z}{\lambda}}+e^{-\frac{z}{\lambda}})}{(e^{\frac{t_c}{2\lambda}}+e^{-\frac{t_c}{2\lambda}})}-1\right)
\end{equation}
For the strain, we find
\begin{equation}
\epsilon_{zz}(t,z)=\frac{\sigma_{zz}}{3K} \frac{(e^{\frac{z}{\lambda}}+e^{-\frac{z}{\lambda}})}{(e^{\frac{t_c}{2\lambda}}+e^{-\frac{t_c}{2\lambda}})}
\end{equation}
Accordingly, we obtain for the displacement component in z-direction 
\begin{equation}
u_{z}(t,z)=\frac{\sigma_{zz}\lambda}{3K} \frac{(e^{\frac{z}{\lambda}}-e^{-\frac{z}{\lambda}})}{(e^{\frac{t_c}{2\lambda}}+e^{-\frac{t_c}{2\lambda}})}, 
\end{equation}
For large $\lambda$, the displacement at the boundary $z=t_c/2$ can be rewritten as 
\begin{equation}
u_{z}(z=t/2)=\frac{\sigma_{zz}}{3K\left(1+\frac{t_c^2}{12\lambda^2}\right)}\frac{t_c}{2}+\mathcal{O}(\frac{1}{\lambda^4}),
\quad 
\end{equation}
where $\mathcal{O}$ denotes the Landau symbol.
Therefore, we may infer an effective compression modulus of the form 
$$K_{eff}=K\left(1+\frac{t_c^2}{12\lambda^2}\right)=\left( K+\frac{ i\omega   \eta t_c^2}{12 k_{perm}}\right).$$
The absolute value of $K_{eff}$ grows with frequency reflecting a trend to approach an effective incompressibility in the large frequency regime.

In the following, we will give a rough order of magnitude estimate of the dissipative term $\frac{ i\omega   \eta t_c^2}{12 k_{perm}}$ in $K_{eff}$ for parameters of the actin cortex layer in mitotic cells.
Based on the Hagen-Poiseuille equation \cite{gran12}, we estimate the permeability of the actin cytoskeleton as $d_{pore}^2/32$, where $d_{pore}$ is the diameter of a cytoskeletal pore which we assume to be $\approx 50\,$nm for the mitotic cortex \cite{char06}.
Furthermore, we estimate the cytoplasmic viscosity $\eta$ inside the cortical pores to be $\approx 10^{-3}$Pa$\cdot$s~ \cite{kalw11}. The length scale $t_c$ is approximated by the previously measured thickness of the cortex ($200\,$nm)~ \cite{clar13}.
We thus obtain an estimate of the dissipative (imaginary) term of $K_{eff}$ of the cortex  of $\approx 100\,$Pa at $f=10\,$Hz. This elastic modulus is still more than an order of magnitude lower than the shear modulus of the mitotic cortex at $10\,$Hz which can be inferred from  \cite{fisc16} to be $\approx 200\,$kPa. Thus, we expect that the dissipative, imaginary term of $K_{eff}$  gives a small, negligible contribution at  frequency $f=10\,$Hz and lower frequencies because  $K\gtrsim G$  (provided that $\nu>0$). 
\hfill

\section{Influence of internal viscosity on cell mechanical response}\label{Sec:viscosity}
\label{sec:AppSec3}
\begin{figure*}[h]
	\centering
	\includegraphics[width=7cm]{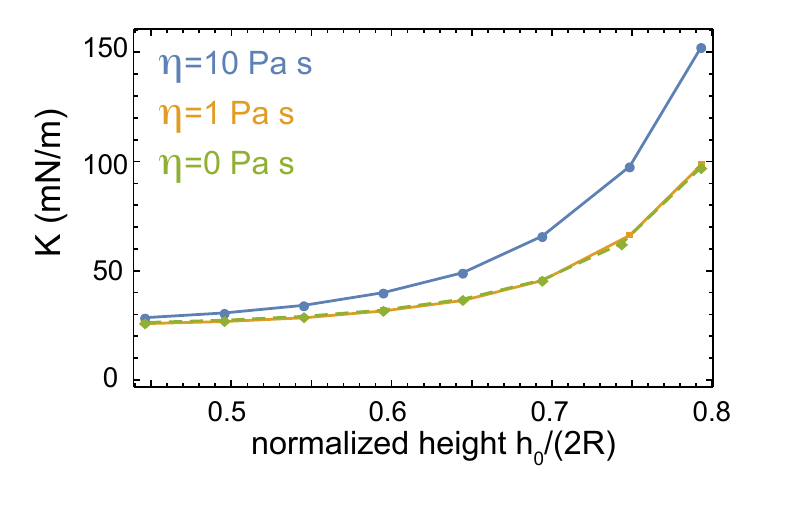}
	\caption{ \label{fig:FigS1}
		Effective shell moduli calculated from simulations in the presence of a viscous bulk of varying viscosity $\eta$. Mechanical parameters of the model cell's shell were $K_B=25\,$mN/m, $K_S=8.33\,$mN/m, $\gamma_a=1.5\,$mN/m, $t=200\,$nm.
		Up to a viscosity of $1\,$Pa$\cdot$s, the influence of the bulk viscosity is negligible. At $\eta=10\,$Pa$\cdot$s, the bulk viscosity increases the calculated effective modulus of the model cell.}
\end{figure*}

In our simulations, we tested the influence of internal cytoplasmic viscosity on the force response of measured cells.
To this end, we simulated the time-periodic deformation of model cells, that were constituted by an elastic shell  with typical cell parameters and a viscous incompressible (pressurized) interior (\autoref{fig:FigS1}). Typical values for the viscosity of the non-cytoskeletal  phase of the cytoplasm range between $10^{-3}-10^{-2}\, {\rm Pa}\cdot{\rm s}$~ \cite{vale05,kalw11}. From our simulations, we find that the force contribution due to viscous friction generated by cyclic cytoplasmic deformation is negligible up to frequencies of $10\,$Hz and  viscosities of $1\, {\rm Pa}\cdot{\rm s}$. There, the calculated effective elastic modulus of the model cell agrees within $1\%$ with the modulus obtained for the case of vanishing internal viscosity (\autoref{fig:FigS1}). This finding suggests that cytoplasmic viscosities give a negligible contribution to the mechanical response of cells during our cell-mechanical probing, which is corroborated by earlier experimental findings \cite{fisc16}.
At a probing frequency of $10\,$Hz, we start to see notable changes of the elastic modulus for $\eta=10\,$Pa$\cdot$s in simulations  (\autoref{fig:FigS1}).

\section{Phenomenological description of the effective elastic modulus}
\label{sec:AppSec7}
We performed simulations of a small uniaxial compression step ($\Delta h=0.5\,\mu$m) of pressurized elastic shells.  Each shell has a thickness of $200\,$nm, cell volume of $4300\,\mu{\rm m^3}$, an area bulk modulus $K_B=25\,$mN/m and an area shear modulus $K_S=8.3, 10, 15, 20$ or $25\,$mN/m. For each value of $K_S$, we simulated compression from an initial reference height of $h_0=15, 14, 13, 12, 11$ or $10\,\mu$m. This was repeated for different values of cortical tension ($\gamma_a=0.5, 1, 1.5, 2, 2.5$ and $3\,$mN/m). Therefore, in total $5\times 6\times 6=180$ simulations have been performed to calibrate the cellular response to uniaxial compression at different mechanical parameters of the cortex. (Furthermore, another $2\times 180$ simulations were performed with equal parameters but deviating bending stiffness testing the influence of changing cortex thickness. There, we assumed i)  the absence of bending rigidity or ii) a twofold increased value of bending rigidity, see Appendix~\ref{sec:AppSec5})\\
From each simulation, we extracted the force exerted on the elastic shell after compression $F_{tot}$ and calculate the effective elastic modulus of the cortical shell $K$ as described in the main text (see Eqn.(1-4), main text).\\
We describe the height dependence of the effective shell elastic modulus as $K(\tilde h)\approx K_B(1+\alpha \exp(\tilde h/\lambda))$ (see Eq.~5, main text).
In this formula, the coefficient $\lambda$ was determined  from an exponential fit of simulated data by the function $K_B(1+\alpha\exp(\tilde h/c))$. 
We noted, that the fit parameter $c$ varied only slightly in dependence of shell parameters $\tilde K_S$ and $\tilde \gamma_a$. To spare the characteristic height scale $c$ as a fit parameter for our noisy experimental data, we used henceforth its average value $\lambda=0.091081$. In turn, we refitted the effective moduli of simulated data by $K_B(1+\alpha\exp(\tilde h/\lambda))$ for set values of $K_B, K_S$ and $\gamma_a$, providing $\alpha$ as a function of the dimensionless parameters $\tilde K_S=K_S/K_B$ and $\tilde \gamma_a=\gamma_a/K_B$. For a given value of $\tilde\gamma_a$, the dependence of $\alpha$ on $\tilde K_S$ is captured by a fit function $C(\tilde\gamma_a) \ln[\tilde K_S] + D(\tilde\gamma_a)$. Finally, the dependence of the fit parameters $C(\tilde\gamma_a)$ and $D(\tilde\gamma_a)$ on the parameter $\tilde \gamma_a$ is captured  through a polynomial fit of third degree:
\begin{align*}
\scriptsize
C_{fit}(\tilde\gamma_a)&=(1.09\cdot10^{-4} + 1.071\cdot 10^{-4}\, \tilde\gamma_a  - 1.54\cdot 10^{-5}\, \tilde\gamma_a^2+ 1.08\cdot 10^{-6} \,\tilde\gamma_a^3),\\
D_{fit}(\tilde\gamma_a)&=(9.418\cdot 10^{-6} -5.874\cdot 10^{-5}\, \tilde\gamma_a - 2.157\cdot 10^{-5}\, \tilde\gamma_a^2+ 4.44\cdot 10^{-6}\, \tilde\gamma_a^3).
\end{align*}
By construction, the resulting function $C_{fit}(\tilde\gamma_a) \ln[\tilde K_S] + D_{fit}(\tilde\gamma_a)$ makes excellent quantitative predictions about the value of $\alpha$ in dependence of $\tilde\gamma_a$ and $\tilde K_S$  (Fig.~2d, main text).\\
For simulations with the alternative assumptions of i) twofold bending stiffness and ii) vanishing bending stiffness,
we obtain different fit polynomials. For i), we have
\begin{align*}
\scriptsize
C_{fit}(\tilde\gamma_a)&=(1.59\cdot10^{-4} + 1.09\cdot 10^{-4}\, \tilde\gamma_a  - 1.15\cdot 10^{-5}\, \tilde\gamma_a^2+ 4.4\cdot 10^{-8} \,\tilde\gamma_a^3),\\
D_{fit}(\tilde\gamma_a)&=(-3.09\cdot 10^{-5} -5.41\cdot 10^{-5}\, \tilde\gamma_a - 3.68\cdot 10^{-5}\, \tilde\gamma_a^2+ 7.87\cdot 10^{-6}\, \tilde\gamma_a^3).
\end{align*}
For ii), we find
\begin{align*}
\scriptsize
C_{fit}(\tilde\gamma_a)&=(4.22\cdot10^{-5} + 1.5\cdot 10^{-4}\, \tilde\gamma_a  - 4.67\cdot 10^{-5}\, \tilde\gamma_a^2+ 6.34\cdot 10^{-6} \,\tilde\gamma_a^3),\\
D_{fit}(\tilde\gamma_a)&=(8.62\cdot 10^{-5} -1.65\cdot 10^{-4}\, \tilde\gamma_a+4.76\cdot 10^{-5}\, \tilde\gamma_a^2- 8.42\cdot 10^{-6}\, \tilde\gamma_a^3).
\end{align*}

\section{Cell deformation simulations}
\label{sec:AppSec8}
Simulations were performed using the finite element (FEM) toolbox AMDiS, developed at the Institute of Scientific Computing TU Dresden \cite{Witkowski2015}. We use an axisymmetric Arbitrary Lagrangian Eulerian (ALE) model with incompressible Navier-Stokes equations for the viscous fluid inside the cell, where the two plates and the forces acting on the membrane are implemented as boundary conditions. 

We assume the cell to be in a stationary state initially, where elastic parameters have no influence on the force exerted by the cell on the plates. The cell is then compressed by a prescribed sinusoidal decrease of the distance $h$ between the plates, while simultaneously calculating the force exerted on the upper plate.

\begin{figure}[!ht]
	\centering
	\includegraphics[width=12cm]{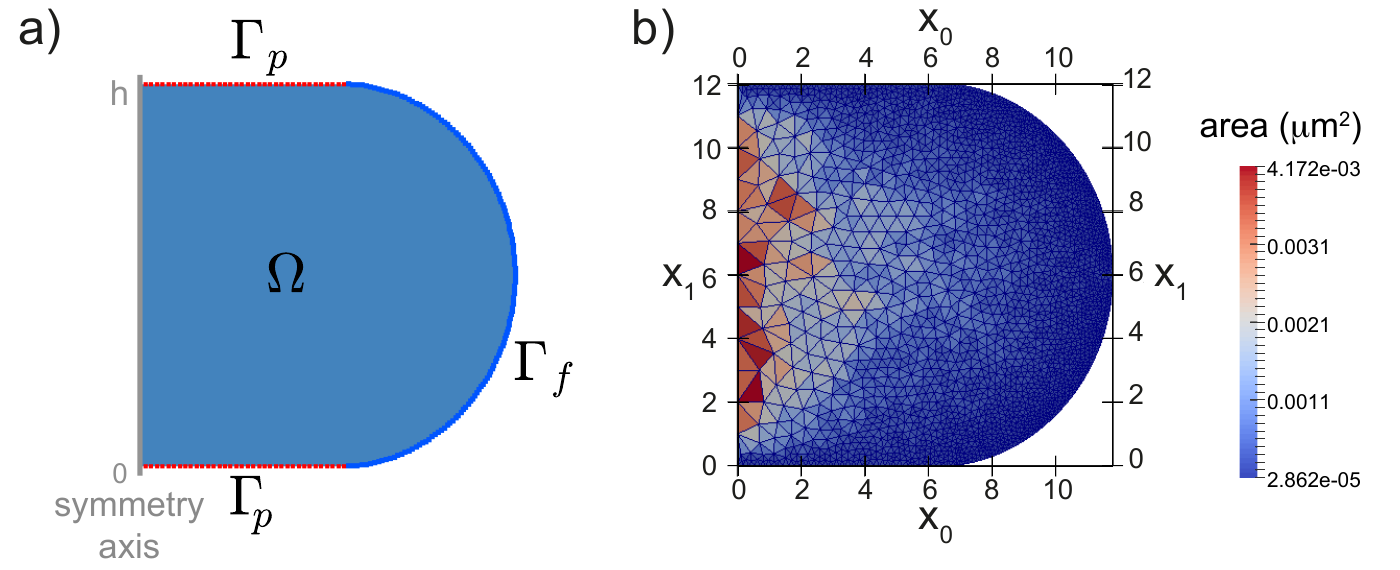}
	\caption[]{\label{fig:mesh} (a) Example image for a simulation domain $\Omega$. The plates here are on the top and bottom sides of the cell with the boundary $\Gamma_p$ (red dotted line). The free part of the boundary of the cell is $\Gamma_f$ (blue line). (b) Example image for the mesh for $h_0=12\,\mu$m. The color scale indicates the areas of the triangles.}
\end{figure}

Using axisymmetry normal to the plates, we can perform  calculations on a two dimensional domain describing half of the cell's cross-section. 
An example image of the simulation domain is shown in \autoref{fig:mesh}. 
The interior of the cell is denoted by the computational domain $\Omega$ which is bounded by the cell cortex/membrane $\Gamma$ and the symmetry axis. $\Gamma$ itself is subdivided into the area touching the plates $\Gamma_p$ and the free surface area $\Gamma_f$. 
During compression a part of the free surface will touch the plate, accordingly $\Gamma_p$ and $\Gamma_f$ are time-dependent:
\begin{align}
\Gamma_p(t) = \left\lbrace \mathbf{x}=(x_0,x_1)\in\Gamma: x_1=0 \lor x_1=h(t) \right\rbrace, \qquad \Gamma_f(t) = \Gamma/\Gamma_p(t).
\end{align} 
The interface curve of $\Gamma$ for the initial meshes with $h=h_0$ is given by a minimal surface calculated according to equations described in \cite{fisc14}. 

The system is governed by the axisymmetric Navier-Stokes equations \cite{Mokbel2018_PF_FSI} together with the surface forces and contact conditions for the plate,
\begin{align}
\rho\,\partial^{\bullet}_t\mathbf{v} &= \nabla\cdot \mathbf{S} +\frac{\eta}{x_0} \begin{pmatrix}2\partial_{x_0}v_0 \\ \partial_{x_0}v_1+\partial_{x_1}v_0 \end{pmatrix} + \frac{2\eta}{(x_0)^2}  \begin{pmatrix} v_0 \\ 0 \end{pmatrix}, && \text{in}\ \Omega \\
\nabla\cdot {\bf v} + \frac{1}{x_0}v_0 &= 0, && \text{in}\ \Omega \\
\mathbf{S} &= \eta\left(\nabla\mathbf{v}+\nabla\mathbf{v}^T\right)-p\mathbf{I}, && \text{in}\ \Omega \\
\mathbf{S}\cdot\mathbf{n} &=- \frac{\partial E_{\text{tension}}}{\partial\Gamma}- \frac{\partial E_{\text{bend}}}{\partial\Gamma}- \frac{\partial E_{\text{stretch}}}{\partial\Gamma}, && \text{on}\ \Gamma_f\\
\left[\mathbf{S}\cdot\mathbf{n}\right]_0 &= \left[- \frac{\partial E_{\text{tension}}}{\partial\Gamma}- \frac{\partial E_{\text{bend}}}{\partial\Gamma}- \frac{\partial E_{\text{stretch}}}{\partial\Gamma}\right]_0, && \text{on}\ \Gamma_p\\
v_1 &= \delta_{x_1>0} \cdot \partial_t h , && \text{on}\ \Gamma_p\\
v_0 &= 0, && \text{on}\ \partial\Omega/\Gamma
\end{align}
where $\partial^{\bullet}_t\mathbf{v}$ denotes the material derivative of the velocity field $\mathbf{v}=(v_0,v_1)$ of the fluid inside the domain, \textit{p} is the pressure, $\eta$ the viscosity, $\mathbf{n}$ the outer unit normal of the domain on $\Gamma$. The first variation of the interfacial energies with respect to changes in $\Gamma$ yields the interfacial forces. We have \cite{Mokbel}
\begin{align}
\frac{\partial E_{\text{tension}}}{\partial\Gamma} = -2\gamma_a H\mathbf{n},\ \frac{\partial E_{\text{bend}}}{\partial\Gamma} = B\left(2\Delta\left(H-H_0\right) + 4\left(2H^2-K_G\right)\left(H-H_0\right)-4H\left(H-H_0\right)^2\right)\mathbf{n}, 
\end{align}
where $\gamma_a$ is the active surface tension, $B$ is the bending stiffness of the cell cortex, $H$ the mean curvature, $H_0$ the mean curvature in the initial state, and $K_G$ the gaussian curvature, respectively. Formulas for the calculation of the curvatures of an axisymmetric surface grid can be found in \cite{HuEtAl2014}.\\
The formula for the elastic force involves the two principal stretches $\lambda_1$ and $\lambda_2$, that describe the relative change of the surface length in lateral and rotational direction, respectively: 
\begin{align}
\lambda_1 = \frac{\text{d}s}{\text{d}s_{|t=0}}\, , \lambda_2 = \frac{x_0}{x_{0|t=0}}\, ,
\end{align} 
where $s$ is the arc length, $x_0$ the distance to the symmetry axis, and $s_{|t=0}$ and $x_{0|t=0}$ are the corresponding quantities at the same material point in the initial state. 
With this, we can write \cite{Mokbel}
\begin{align}
\frac{\partial E_{\text{stretch}}}{\partial\Gamma} = \left(2H\,\textbf{n} -\nabla_{\Gamma}\right)\left[\left(K_B+K_S\right)\left(\lambda_1 - 1\right) + \left(K_B-K_S\right)\left(\lambda_2 - 1\right)\right] - 2K_S\left(\lambda_1-\lambda_2\right)\,\frac{1}{x_0}\begin{pmatrix}
1 \\ 0
\end{pmatrix}.
\label{formula:elasticforce}
\end{align}
The discretization is done by an ALE method, where grid points at the cell surface $\Gamma$ are moved with the velocity ${\bf v}$. 
Interior grid points in $\Omega$ are displaced by a harmonic field $\mathbf{w}$ calculated in every time step:
\begin{align}
\Delta \mathbf{w} &= 0, && \text{in}\ \Omega\nonumber\\
\mathbf{w} &= \tau\mathbf{v}, && \text{on}\ \partial\Omega\label{formula:law}
\end{align}
where $\tau$  is the time step size.
Whenever a grid point of the free boundary, $\mathbf{x}=(x_0,x_1)\in\Gamma_f$, reaches the lower or upper plate, $x_1\leq 0$ or $x_1\geq h$, it is moved exactly onto the plate, i.e. $x_1=0$ or $x_1=h$, respectively, and we mark the point as a member of the discrete points set of $\Gamma_p$ instead of $\Gamma_f$. 

The position (and velocity) of the moving plate are prescribed by a cosine function
\begin{align}
h(t) &= h_0-\frac{\Delta h}{2}\left[1-\cos\left(f\cdot(t-t_0)\right)\right],
\end{align}
where $f$ is the oscillation frequency. The compression starts at $t=t_0$ and ends at $\tilde{t}=\pi/ f+t_0$, where maximum compression is reached. For $t>\tilde{t}$, we keep the cell in the compressed state, $h(t)=h_0-\Delta h$.

The force, exerted on the right plate, is calculated in every time step using
\begin{align}
F_{\text{plate}} = 2\pi \int_{\Gamma} a(x_0,x_1)\cdot x_0\cdot \left[\mathbf{S}\cdot\mathbf{n}\right]_1\ \text{d}\Gamma\, .
\end{align}
where a$:\Gamma\rightarrow\mathbb{R}$ is the piecewise linear extension of an indicator function for the right (upper) plate: 
\begin{align}
a(x_0,x_1) = \begin{cases}
1 &\quad\text{if} \quad x_1\geq h\\
0 &\quad\text{else} \, .
\end{cases}
\end{align}
After compression, i.e. at height $h_1$, we have $F_{plate}=F_{tot}$, cf. Eq.~(2) main text. 

As shown in  see \autoref{Sec:viscosity}, we found that the contribution of interior viscosity to the force response is negligible. 
To simulate the process without interior viscosity, one can take advantage of some simplifications. In this case, we do not need a 2D mesh representing half of the cell's cross section but only a 1D mesh representing the membrane $\Gamma$ in \autoref{fig:mesh}(a). Then, in every timestep we calculate the displacement of the membrane points according to 
\begin{align}
\textbf{v} &= \mu \, \left[- \frac{\partial E_{\text{tension}}}{\partial\Gamma}- \frac{\partial E_{\text{bend}}}{\partial\Gamma}- \frac{\partial E_{\text{stretch}}}{\partial\Gamma}-p_c(V-V_0)\,\mathbf{n}\right]\, ,&&\text{on}\ \Gamma_f \\
\mathbf{v} &= \begin{pmatrix}
\mu\,\left[- \frac{\partial E_{\text{tension}}}{\partial\Gamma}- \frac{\partial E_{\text{bend}}}{\partial\Gamma}- \frac{\partial E_{\text{stretch}}}{\partial\Gamma}\right]_0 \\\delta_{x_1>0}\cdot\partial_t h 
\end{pmatrix}\, ,&&\text{on}\ \Gamma_p 
\label{formula:law1D}
\end{align}
where $\mu$ is the (inverse)  coefficient of friction, here $\mu=1.25\cdot 10^{-7}\,$m$^2$s/kg, $V$ is the (3D) volume of the cell, $V_0$ is the volume in the initial state and $p_c$ is a large constant to provide the pressure to ensure volume conservation, here $p_c = 1.2\cdot 10^5\,$N/m$^5$.

The interfacial forces are implemented explicitly, i.e. curvatures and principal stretches of the configuration in the previous time step are used to calculate the force in the new time step. Therefore, the system is quite restrictive to time step sizes. For the simulations with interior flow, a time step of $0.5\mu$s is used. For a period frequency of $1\,$s we need $255.000$ time steps until the end time of $0.51\,$s is reached. The initial mesh is shown in \autoref{fig:mesh}(b) for $h_0=12\,\mu$m. A fine mesh at the membrane is necessary to produce highly accurate results for the membrane forces. Hence, the triangle sizes amount from approximately $0.003\,\mu$m$^2$ at the interface to $0.4\,\mu$m$^2$ around the cell center.

\section{Materials and Methods}
{\bf Cell culture.}
We cultured  HeLa Kyoto cells expressing a green-fluorescent histone construct  (H2B-GFP) and red-fluorescent membrane label  (mCherry-CAAX)  in DMEM (PN:31966-021, life technologies) supplemented with 10\% (vol/vol)~fetal bovine serum, $100\,\mu$g/ml penicillin, $100\,\mu$g/ml streptomycin and $0.5\,\mu$g/ml geneticin (all Invitrogen) at 37\tss{\circ}C with 5\% CO$_{2}$. 
One day prior to the measurement, 10000 cells were seeded into a silicon cultivation chamber (0.56~cm$^2$, from ibidi 12 well chamber )  that was placed in a 35~mm cell culture dish (fluorodish FD35-100, glass bottom) such that a confluency of $\approx 30\%$ is reached at the day of measurement.
For AFM experiments, medium was changed to DMEM (PN:12800-017, Invitrogen) with 4~mM NaHCO$_{3}$  buffered with 20~mM HEPES/NaOH pH~7.2. 
Mitotic  arrest of cells was achieved by addition of  S-trityl-L-cysteine (STC, Sigma) two to eight hours before the experiment at a concentration of $2\,\mu$M. This allowed conservation of cell mechanical properties  during measurement times of up to $30\,$min for one cell \cite{skou06}.
Cells in mitotic arrest were identified by their shape and/or H2B-GFP. Diameters  of uncompressed, roundish, mitotic cells typically ranged from $19-23\,\mu$m.

{\bf Atomic Force Microscopy.}
The experimental set-up consisted of an AFM (Nanowizard I, JPK Instruments) mounted on a Zeiss Axiovert 200M optical, wide-field microscope. For imaging, we used  a 20x objective (Zeiss, Plan Apochromat, NA=0.80) and a CCD camera (DMK 23U445 from theimagingsource). During measurements, cell culture dishes were kept in a petri dish heater (JPK instruments) at 37$^\circ$C. On every measurement day, the spring constant of the cantilever was calibrated using the thermal noise analysis (built-in software, JPK). 	
Cantilevers were tipless, $200-350\,\mu$m long, $35\,\mu$m wide, $2\,\mu$m thick
(NSC12/tipless/noAl or CSC37/tipless/noAl, Mikromasch) with nominal force constants between $0.3$ and $0.8\,$N/m.
Cantilevers were modified with wedges to  correct for the $10^\circ$ cantilever tilt  consisting of UV curing adhesive  (Norland 63) \cite{stew13} . 
During measurements, measured force, piezo height and time were output at a time resolution of  $100\,$Hz.

{\bf Cell compression protocol.}
Prior to cell compression, the AFM cantilever was lowered to the dish bottom near the cell until it came into contact with
the surface and then retracted to $\approx 15\,\mu$m above the surface. Thereafter, the free cantilever was moved over the cell. At this stage, a brightfield picture of the equatorial plane of the confined cell is recorded to estimate the area of the equatorial cross-section and in turn to estimate cell volume as described in \cite{fisc16}.
The cantilever was then gradually lowered in steps of  $0.5$ or $1\,\mu$m at a set speed of $0.2\,\mu$m/s interrupted by waiting times of $50-150\,$s. During this waiting time, we performed  sinusoidal oscillations around the mean cantilever height at different frequencies ($f=0.02, 0.1,1$ and $10$~Hz) with a piezo height amplitude of $0.25\,\mu$m. The cycle of compression and subsequent oscillations around a constant mean height was repeated until the cell started to bleb which was typically at a height of $10\,\mu$m.
For frequencies $f=0.1-10$~Hz, height oscillation were performed  for $\geq 5$ periods.
For frequency $f=0.02\,$Hz , height oscillation were performed  for $\geq 2$ periods. 
For a first subset of cells ($N\approx 50$), mechanical probing was performed jointly at frequencies $f=0.1, 1, 10$~Hz, for a second subset of cells ($N\approx 10$), all frequencies ($f=0.02, 0.1, 1, 10$~Hz) were measured on one cell, for a third subset  of cells ($N\approx 25$), only the slow frequency of $f=0.02\,$Hz was measured, in order to limit the overall measurement time on one cell.
During the entire experiment, the force acting on the cantilever was continuously recorded. The height of the confined cell was computed as the difference between the height that the cantilever was raised from the dish surface and lowered onto the cell plus the height of spikes at the rim of the wedge (due to imperfections in the manufacturing process \cite{stew13}) and the force induced deflection of the cantilever. We estimate a total error of cell height of $\approx 0.5\,\mu$m due to unevenness of the cantilever wedge and due to  vertical movement of the cantilever to a position above the cell. 

{\bf Data analysis.}
Geometrical parameters of each analyzed  cell (such as contact area $A_c$ with the wedge, mean curvature $H$ of the free cell surface and cell surface area $A$) are for each cell estimated as previously described in \cite{fisc16}. In turn, these parameters are used to calculated the effective cortical tension $\gamma_{eff}$ according to Eq.~2, main text.\\
Since we impose only small deformation oscillations on the cell, we may use an analysis scheme in the framework of linear viscoelasticity, as shown in our previous work \cite{fisc16}, Supplementary.
Oscillation amplitudes of  effective cortical tension $\hat \gamma_{eff}$ and cell surface area $\hat A$ where determined by performing a linear fit using the fit function $a\cos(2\pi t/T)+b\sin(2\pi t/T)+c t$ where $T$ is the oscillation period of the  imposed cantilever oscillations. The oscillation amplitude was then calculated as $a^2+b^2$. 
The strain amplitude $\hat \epsilon_A$ was calculated as $\hat A/<A>$.\\
For data analysis, only cells were considered that had a roughly constant average cortical tension during the measurement (not more than $10\%$ deviation). This was true for $\approx 70\%$ of the cells. Major variations in the cortical tension could mostly be attributed to visible blebbing events. \\
For the calculation of cortical Poisson ratios, we demanded that oscillatory measurements of cells had to be in a range of normalized height $\tilde h$ between $0.5$ and $0.75$ to match the parameters of the simulations.  Only cells with at least four different heights sampled in this range were considered for analysis, where the highest normalized height had to be larger than $0.68$.
Furthermore, we demanded that the r-squared value of the exponential fit of the obtained effective elastic modulus according to of Eq.~5, main text, had to be larger than 0.5. This constraint was released for Poisson ratio estimates larger than 0.7 since this indicates an almost constant value of effective modulus in dependence of cell height. For the case of a constant functional dependence, the fit cannot be better than the approximation of the data by the mean, leading to an r-squared value that approaches zero.

\section{Distribution of estimated Poisson ratios at different frequencies}
\label{sec:AppSec4}
\begin{figure*}[b]
	\centering
	\includegraphics[width=9cm]{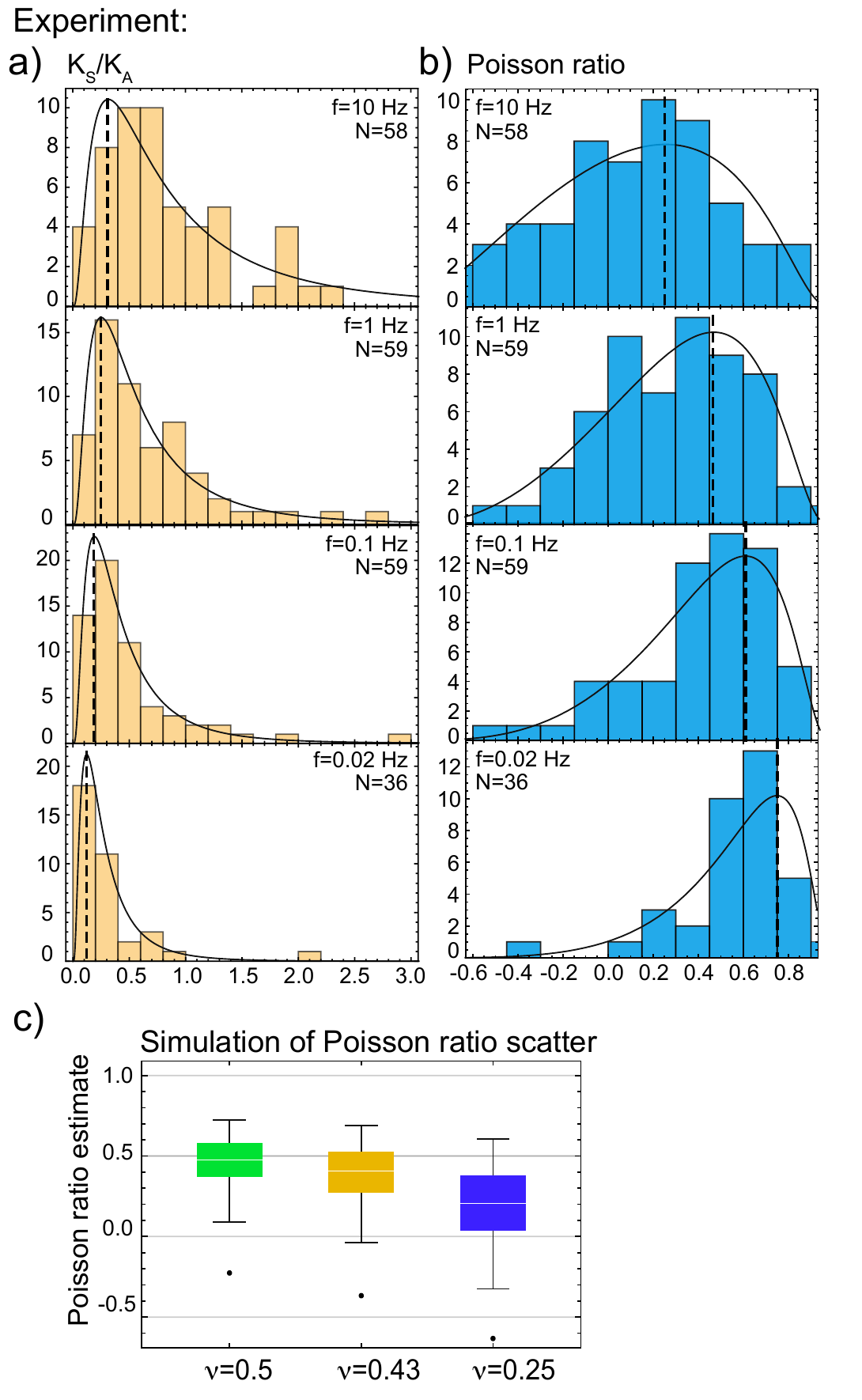}\label{fig:FigS2}
	\caption{
		a,b) Results of the analysis of experimental data.
		a) Histograms of estimated values for the ratio between area shear modulus and area bulk modulus $\tilde K_S=K_S/K_B$ with fitted lognormal distributions (logarithmic mean and standard deviation: f=10 Hz, $\mu=-0.31,\sigma=0.92$; f=1 Hz, $\mu=-0.7,\sigma=0.82$; f=0.1 Hz, $\mu=-1.03,\sigma=0.8$; f=0.02 Hz, $\mu=-1.5,\sigma=0.8$).
		b) Histograms of estimated Poisson ratios with induced distributions.
		c) Poisson ratio estimates for simulated data including fake errors of cell volume and cell height. Green data set, $\nu=0.5$: median: 0.48, IQR: 0.21; orange data set, $\nu=0.43$: median: 0.41, IQR: 0.25; blue data set, $\nu=0.25$: median: 0.20, IQR: 0.34, where IQR is the interquartile range, i.e. the distance between 25th and 75th percentile. The sample size was N=55, each, and therefore similar to our experimental sample sizes.}
\end{figure*}

In \autoref{fig:FigS2}(a,b), we present the histograms of measured values of the ratio between area shear modulus and area bulk modulus $\tilde K_S=K_S/K_B$ and 
associated Poisson ratios.  Histograms of $\tilde K_S$ have been fitted with the lognormal distribution of maximum likelihood. Histograms of Poisson ratio $\nu$ are plotted with the distribution induced by the lognormal distribution of $\tilde K_S$. This induced distribution is calculated by the functional relationship $\nu=(1-\tilde K_S)/(1+\tilde K_S)$.\\
In \autoref{fig:FigS2}(c), we analyzed simulated data in the same way as experimental data, however not using the exact values for cell volume and cell height. Instead, we drew the values of cell volume and cell height from a Gaussian distribution with correct mean value and a standard deviation that matches our error estimate for cell volume and cell height ($7.5\%$ and $0.5\,\mu$m, respectively). We chose $N=55$, similar to sample numbers measured in the experiments. The resulting scatter in estimated Poisson ratios is shown in \autoref{fig:FigS2}(c), where the horizontal lines indicate median, 25th percentile and 75th percentile. While the median is close to actual values of the Poisson ratio,  the resulting scatter is substantial, in particular for $\nu=0.25$. We conclude that the large scatter of Poisson ratios observed in our experimental data does not exclusively result from cell-cell variations but stems to a substantial amount from experimental uncertainties.

\section{Influence of cortical thickness and bending stiffness variations on Poisson ratio estimates}\label{sec:influence of bending}
\label{sec:AppSec5}
\begin{figure}[h]
	\centering
	\includegraphics[width=\textwidth]{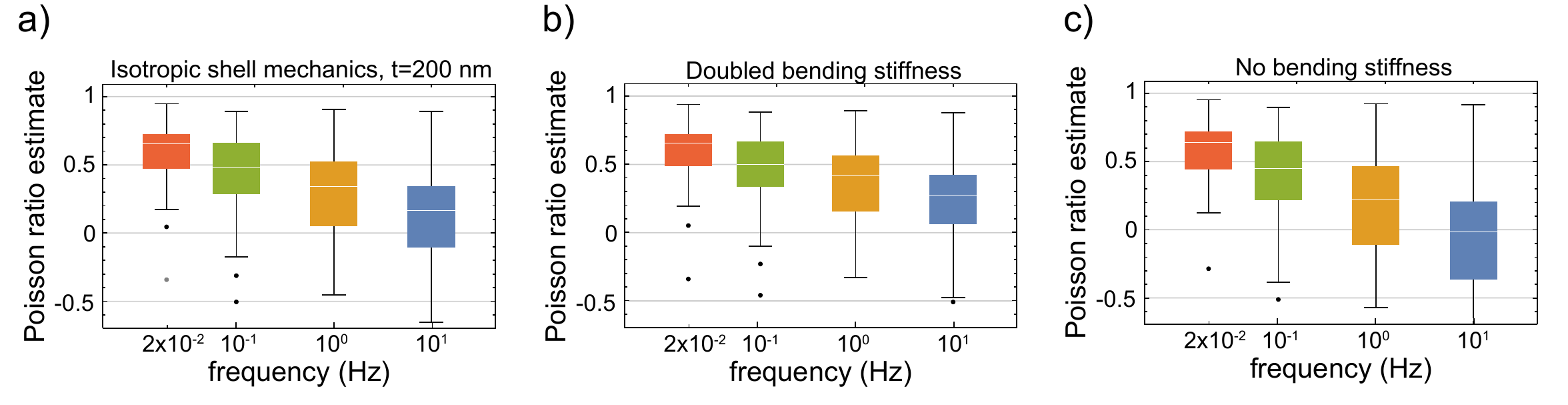}\label{fig:FigS3}
	\caption{Poisson ratio estimates of the mitotic cortex of HeLa cells using different assumptions on cortical bending stiffness in our analysis (sample sizes from left to right: N=36, N=59, N=59, N=58).
		a) Estimated Poisson ratios using the assumption of an isotropic cortex with thickness of $200$ nm as printed in Fig. 3f, main text. From left to right, median values: 0.66, 0.48, 0.34, 0.17, IQR: 0.26, 0.38, 0.48, 0.45.
		b) Estimated Poisson ratios using twofold bending stiffness as compared to the analysis in the main text. From left to right, median values: 0.65, 0.5, 0.41, 0.28, IQR: 0.24, 0.33, 0.41, 0.36.
		c) Estimated Poisson ratios using vanishing bending stiffness. From left to right, median values: 0.64, 0.45, 0.22, -0.013, IQR: 0.3, 0.43, 0.58, 0.57.
	}
\end{figure}

Cortical thickness in mitotic HeLa cells has previously been  estimated to be  $\approx 200\,$nm \cite{clar13}. However, cells exhibit cell-cell variations in cortical thickness and thus variations in cortical bending stiffness which will contribute to scatter of Poisson ratio estimates in our analysis.
In order to examine the influence of cell-cell variations in cortical bending stiffness, we repeated our simulations of cell deformation with i) twofold increased bending stiffness (corresponding to $\approx 40\%$ relative increase in cortex stiffness) and ii) vanishing bending stiffness of the cortical shell.
Using these alternative simulations to calibrate cell mechanical response, 
we reanalyzed our data. Corresponding alternative Poisson ratio estimates are presented in \autoref{fig:FigS3}(b) and (c) where results of the original analysis from Fig. 4(c), main text, are depicted again in \autoref{fig:FigS3}(a) for direct comparison. 
We see that i) the assumption of higher bending stiffnesses of the cortex would lead to consistently higher Poisson ratio estimates for cortical shells. Furthermore, we see that ii) assuming vanishing bending stiffness would consistently lead to lower Poisson ratio estimates of cortical shells. In both cases, the change in Poisson ratio estimates is particularly striking for Poisson ratio values substantially below 0.5. 
In summary, we conclude that i) an underestimation of cortical bending stiffness in our analysis of experimental data would lead to a consistent underestimation  of cortical Poisson ratios, while ii) an overestimation of cortical bending stiffness in our analysis would lead to a consistent overestimation of cortical Poisson ratio in particular if cortical Poisson ratio values are substantially below 0.5.
Finally, independent of a possible under- or overestimation of absolute values of the Poisson ratio, we find in all cases a significant trend of Poisson ratio increase with decreasing frequency.

\section{The two-dimensional Poisson ratio of a thin shell}
\label{sec:AppSec2}
For the surface energy density of  stretching of a thin shell, one obtains in complete analogy to the three-dimensional case \cite{land86}
\begin{equation}
F=K_S (\epsilon_{ij}-\frac{\delta_{ij}}{2}\epsilon_{ll})^2+\frac{1}{2}K_B (\epsilon_{ll})^2,
\end{equation}
where $K_B$ and $K_S$ are the area bulk modulus and the area shear modulus of the shell and $\epsilon_{ij}, (i,j=1,2)$ are  in-plane coefficients of the strain tensor. 
The corresponding stress-strain relationship reads
\begin{equation}
\sigma_{ij}=2K_S (\epsilon_{ij}-\frac{\delta_{ij}}{2}\epsilon_{ll}) + K_B \delta_{ij}\epsilon_{ll}.
\end{equation}
Correspondingly, the strain can be written as
\begin{equation}
\epsilon_{ij}=\frac{1}{2K_S} (\sigma_{ij}-\frac{\delta_{ij}}{2}\sigma_{ll}) + \frac{1}{4 K_B} \delta_{ij}\sigma_{ll}.
\label{eq:strainEq}
\end{equation}
In the following, we will determine the expression of the two-dimensional Poisson ratio of a thin shell as a function of $K_B$ and $K_S$. To this end, we consider the special case of  a thin, flat, square-shaped patch of a shell subject to a uniform in-plane stretch through opposite forces acting at the top and bottom edge and with free side edges. Correspondingly, the only non-vanishing stress component is $\sigma_{yy}$. 
According to \autoref{eq:strainEq}, the strain tensor is given as
\begin{equation} \label{eq:eps}
\boldsymbol{\epsilon}=
\frac{1}{2K_S}\begin{pmatrix}
\frac{-\sigma_{yy}}{2} &0 \\
0 & \frac{\sigma_{yy}}{2}
\end{pmatrix}
+\frac{1}{4K_B} 
\begin{pmatrix}
\sigma_{yy} &0 \\
0 & \sigma_{yy}
\end{pmatrix}.
\end{equation}
The two-dimensional Poisson ratio $\nu_{2D}$ is defined as the ratio $\nu_{2D}=-\epsilon_{xx}/\epsilon_{yy}$. With the above relation \eqref{eq:eps}, this equates to $\nu_{2D}=(K_B-K_S)/(K_B+K_S)$. Using the definitions $K_B=t E/(2(1-\nu))$ and $K_S=t E/(2(1+\nu))$ for an isotropic shell material with Young's modulus $E$ and Poisson ratio $\nu$, we obtain $\nu_{2D}=\nu$. If the constraint of isotropy is released, the two-dimensional Poisson ratio may adopt values in the range $[-1, 1]$.


\end{document}